\documentclass[aps,prc,reprint,onecolumn,notitlepage,nofootinbib]{revtex4-1}
\usepackage{texnames}
\usepackage[T1]{fontenc}
\usepackage{graphicx,color}
\usepackage{amsmath}
\usepackage{multirow}
\newcommand{\beq}{\begin{equation}}
\newcommand{\eeq}{\end{equation}}
\newcommand{\beqn}{\begin{eqnarray}}
\newcommand{\eeqn}{\end{eqnarray}}

\begin{document}
\title{Testing two-nucleon transfer reaction  mechanism with elementary modes
of excitation in exotic nuclei}

\author{R.A. Broglia$^{1,2}$, G.Potel$^{3,4}$, A. Idini${^5}$, F. Barranco$^6$ and E. Vigezzi$^7$ }
\affiliation{$^1$Dipartimento di Fisica, Universit\`a di Milano,
Via Celoria 16, 
I-20133 Milano, Italy}
\affiliation{$^2$The Niels Bohr Institute, University of Copenhagen, 
DK-2100 Copenhagen, Denmark}
\affiliation{$^3$ National Superconducting Cyclotron Laboratory, Michigan State University, East Lansing, Michigan 48824, USA}
\affiliation{$^4$ Lawrence Livermore National Laboratory L-414, Livermore, CA 94551, USA}
\affiliation{$^5$ Department of Physics, University of Jyvaskyla, FI-40014 Jyvaskyla, Finland}
%\author{F. Barranco}
\affiliation{$^6$ Departamento de F\'isica Aplicada III,
Escuela Superior de Ingenieros, Universidad de Sevilla, Camino de los Descubrimientos, 	Sevilla, Spain}	
%\author{E. Vigezzi}
\affiliation{$^7$INFN Sezione di Milano, Via Celoria 16, I-20133 Milano, Italy }

\begin{abstract}
Nuclear Field Theory of structure and reactions  is confronted with 
observations made on neutron halo dripline nuclei, resulting in the  prediction
of a novel (symbiotic) mode of nuclear excitation, and on the observation of the virtual effect of the halo
phenomenon in the apparently non-halo nucleus $^7$Li. This effect is forced to become real by intervening 
the virtual process with an external (t,p) field which, combined with accurate predictive abilities concerning  the absolute 
differential cross section, reveals an increase of a factor 2 in the  cross section  due to the presence 
of halo ground state correlations, and is essential to reproduce the value of the observed $d \sigma(^7$Li(t,p)$^9$Li)/d$\Omega$.
% revealing an unexpected and stringent test  of the Axel-Brink hypothesis.
%A complete characterization of the structure of nuclei can be obtained by combining 
%information arising from inelastic scattering, Coulomb excitation and $\gamma-$decay, together
%with one- and two-particle transfer reactions. In this way it is possible to probe 
%the single-particle and collective components of the nuclear many-body wavefunction
%resulting from their mutual coupling and diagonalising the low-energy Hamiltonian. 
%We address the question of how accurately such a description can account for experimental
%observations.  It is  concluded that renormalizing empirically and on equal footing  bare single-particle 
%and collective  motion in terms of self-energy (mass) and vertex corrections (screening), 
%as well as  particle-hole and pairing  interactions through particle-vibration 
%coupling allows theory to provide  an overall, quantitative account of the data. 
%The use of the SkM* interaction then leads to a good agreement between theoretical  and experimental lifetimes.
\end{abstract}
\maketitle

\section{Foreword}

At the basis of single-particle motion, fermionic elementary modes of nuclear excitation,  one finds delocalization, measured by the quantality parameter
($q\ll1$ localisation, $q\sim1$ delocalization \cite{Mottelson}), ratio   of the  kinetic energy (ZPF) 
of confinement, and of the  strength of the NN-interaction ($V_0 =$- 100 MeV, $a \approx$  1 fm),
\begin{equation}
q= \frac{\hbar^2}{ma^2} \frac{1}{|V_0|}  \approx 0.4.
\end{equation}

At the basis of BCS pairing  one finds Cooper pairs and independent pair motion, in which the partner nucleons are correlated over distances of the order of 
\begin{equation}
\xi = \frac{\hbar v_F}{2 E_{corr}} \approx  20 {\rm  fm}
\end{equation}
in keeping with the value of $E_{corr} \approx 1-1.5$ MeV (see e.g. \cite{Brogliaetal1971c}) displayed by pair addition and pair subtraction modes \cite{Bohr1964,BohrMottelsonI,BesandBroglia1966}
around closed shell nuclei 
($E_{corr} \approx \Delta$ in superfluid systems  ($\approx 1.5$ MeV in $^{120}$Sn) \cite{BohrMottelsonI,BohrMottelsonII}), and the fact that, for nuclei along the stability valley,  $v_F/c \approx$  0.3. 
The (generalised) quantality parameter associated with Cooper pairs can be redefined as 
\begin{equation}
q' = \frac{\hbar^2}{2 m \xi^2} \frac{1}{2E_{corr}} \approx 0.02,
\end{equation}
implying localization. In other words, in a Cooper pair, each nucleon  is solidly anchored  to its partner  leading to an emergent property: rigidity  in gauge 
space. In keeping with the fact that the Cooper pair transfer  cross section $\sigma \sim \sum_{\nu>0} U_{\nu} V_{\nu} = (\Delta/G)^2 \sim (N(0))^2$ is proportional to the square of  
of  the density of levels $N(0)$, Cooper tunneling takes place essentially as successive transfer (without breaking the pair) as a particle of mass $2m$ which sets instantaneously into rotation (vibration) 
superfluid (normal) nuclei, in gauge space \cite{BesandBroglia1966}.
Adding to independent particle motion and pair addition and subtraction modes correlated particle-hole vibrations, complete the elementary modes
of excitation count \cite{Bohr1964} around closed shell nuclei. This  basis of states  is able to provide a first overall picture
 of the low energy spectra as probed by nuclear reactions.

However, the basis  is non-orthogonal and violates Pauli principle, in keeping with the fact that all
 the degrees of freedom  of the nucleus are exhausted by the nucleonic degrees of freedom. 
Pauli exchanging and  orthogonalizing  it with the help of NFT rules \cite{Bes1976a}-\cite{Bortignonetal1977}, together with   two-nucleon transfer reaction theory  
 (second order DWBA describing  simultaneous and successive transfer corrected for non-orthogonality, see refs. \cite{Broglia1999,Broglia1973,Potel2013}  and refs. therein),
 one can calculate the variety of absolute cross sections and transition probabilities which can be directly  compared with the experimental data. 

\section{Pairing vibrations of N=6  magic number isotopes}

As a result  of the (mainly quadrupole) dressing and Pauli exchange of the $2s_{1/2}$ and of the $1p_{1/2}$ orbitals respectively \cite{Sagawa,EPJ},
parity inversion takes place in an island of light nuclei at the drip line. As a result,  the $N=8$ closed shell dissolves,
$N=6$ becoming a novel magic number. This has profound effects in the associated (multipole) pairing vibrational spectrum. 
In particular for  $^9_3$Li$_6$, in which case one is confronted with exotic monopole and  dipole pair addition modes 
($|^{11}$Li(gs)$>$, 
$|^{11}$Li$(1^-; 0.4 {\rm MeV})>$ namely  the Giant Dipole Pygmy Resonance (GDPR) and with an, apparently, 
normal pair removal  mode ($|^7$Li(gs)$>$). 
At the basis of the almost degenerate $0^+$ and $1^-$ pair addition modes one finds 
the fact that in $^{10}$Li (not bound) the $s_{1/2} $ and $p_{1/2}$ orbitals are both at threshold lying close in energy 
($\epsilon_{s_{1/2}} \approx$  0.2 MeV, $\epsilon_{p1/2} \approx  0.5 $ MeV). 
They are thus not available to contribute to standard nuclear Cooper pairing ($^1S_0$ short range NN-potential) .
Induced pairing becomes overwhelming. In keeping with the heavily dressed inverted pairing $s,p$ orbitals, the 
GDPR mode ($E_x \leq 1 $ MeV,  $\approx 8\% $ of TRK) exchanged between $s^2_{1/2}(0)$ and $p^2_{1/2}(0)$
configurations provides most of the glue binding the halo neutron Cooper pair to the core $^9$Li \cite{EPJ}, 
as testified by $^1$H($^{11}$Li,$^9$Li(gs))$^3$H absolute 
cross section.  The population of the first excited state of $^9$Li  ($^1$H($^{11}$Li,$^9$Li($1/2^-$))$^3$H)
provides information on phonon induced pairing  mechanism \cite{Tanihata2008,Potel2010,Potel2014}. 
This is the reason
why the pair of symbiotic states under discussion are boxed in Fig. 1. They are expected  to be a new (composite)
elementary mode of excitation.  

Turning back to the probing of this $1^-$  mode, it could be illuminating in
shedding light into its actual   structure (low energy E1-vortex-like mode, i.e. 
a Cooper pair with angular momentum and parity  $1^-$), to carry out the $^9$Li(t,p)$^{11}$Li reaction. Aside from weak $Q-$value
effects and simple geometrical factors, one will be able to relate the "intrinsic" contribution to the absolute cross sections associated with the population of the ground
state   and of the 1$^-$ state
\footnote{These two states  paradigmatically represent  the
competition between paired and aligned coupling schemes, which play such an important role in defining e.g.   quadrupole shape transitions (see ref. \cite{BohrMottelsonII} and
refs. therein). } . One can then test the role ground state correlations  ({\it gsc}) play in both states.To the extent that the $1^-$ state can be viewed as 
a particle-hole-like (2qp) excitation, {\it gsc} will  decrease the cross section, the inverse  being expected to be the case if this state is the dipole pair addition mode of $^9$Li
(vortex-like Cooper pair). These effects should be reversed concerning the intensity of the 
$\gamma-$decay, as discussed in \cite{Broglia1971a}. How these relations  get qualified in the case of the exotic system under discussion is an open question, which  may benefit from the analogies to be drawn concerning the situation encountered  in connection with the first $0^{+*}$   excited state of $^{12}$Be. 

In fact,  it is  posited that the pair of $0^{+*},  1^-$ (boxed) states of $^{12}$Be displayed in Fig. 1, are 
(part of ?)  the corresponding symbiotic states of $^{11}$Li, modified by the extra binding energy provided 
by the fourth proton. In this case, the possibility of  studying  the new proposed elementary mode of excitation
with a variety of  probes  is richer, due  to the greater stability of the $|^{12}$Be$(0^{+*}$, 2.24  MeV)$>$  state as
compared to the |$^{11}$Li(gs)$>$ .

\begin{figure}
	\begin{center}
\includegraphics[width=\textwidth]{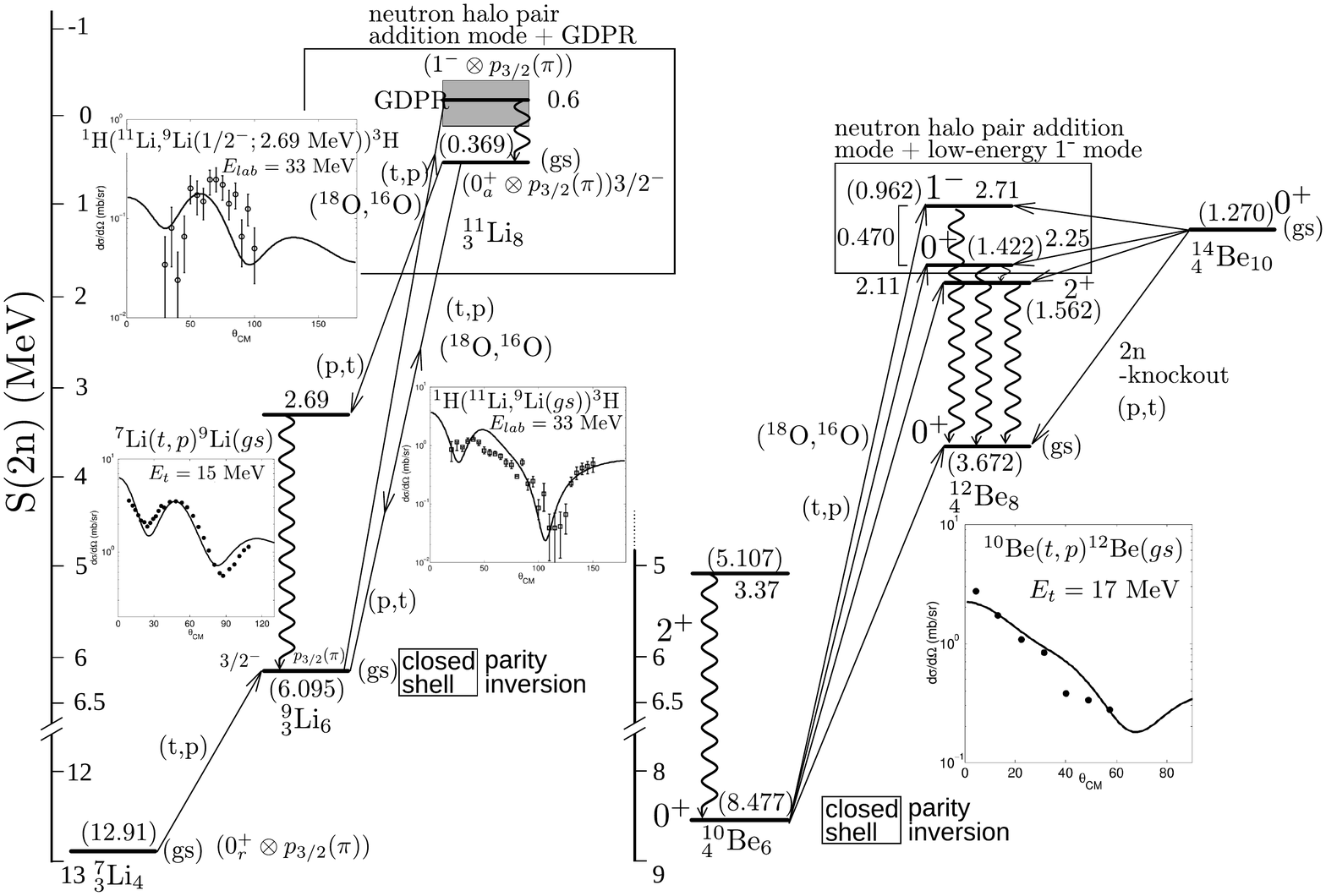}
\caption{ Monopole pairing vibrational modes associated with 
$N=6$ parity inverted closed shell isotopes, together with low-energy E1-strength modes. 
The levels are  displayed as a function of the two-neutron separation energies $S(2n)$. 
These quantities are shown in parenthesis on each level, the excitation energies with respect to the ground state are quoted in MeV. 
Absolute differential cross sections from selected (t,p) and (p,t) reactions calculated as described in the text (cf. \cite{Potel2010,Potel2014}), 
in comparison with the experimental data \cite{Young1971,Fortune1994}.}
\end{center}
\end{figure}

 \begin{figure}
 \begin{center}
\includegraphics[width=0.8\textwidth]{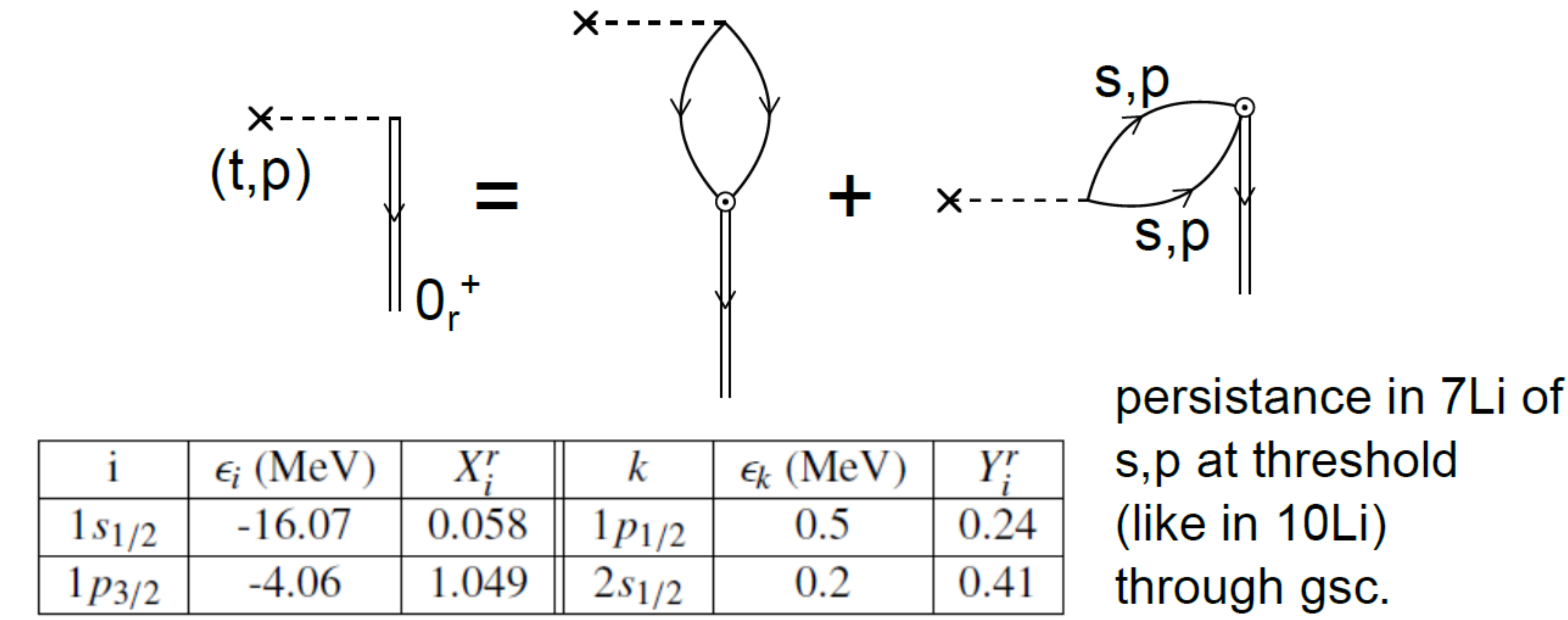}
\caption{ RPA wavefunction of  the pair removal mode ($|gs(^7Li)>$) 
of the closed shell $N=6$ parity  inverted system $^9$Li obtained 
solving the dispersion relation graphically displayed in the upper part of the figure.}
\end{center}
\end{figure}

 \begin{figure}
 \begin{center}
\includegraphics[width=0.8\textwidth]{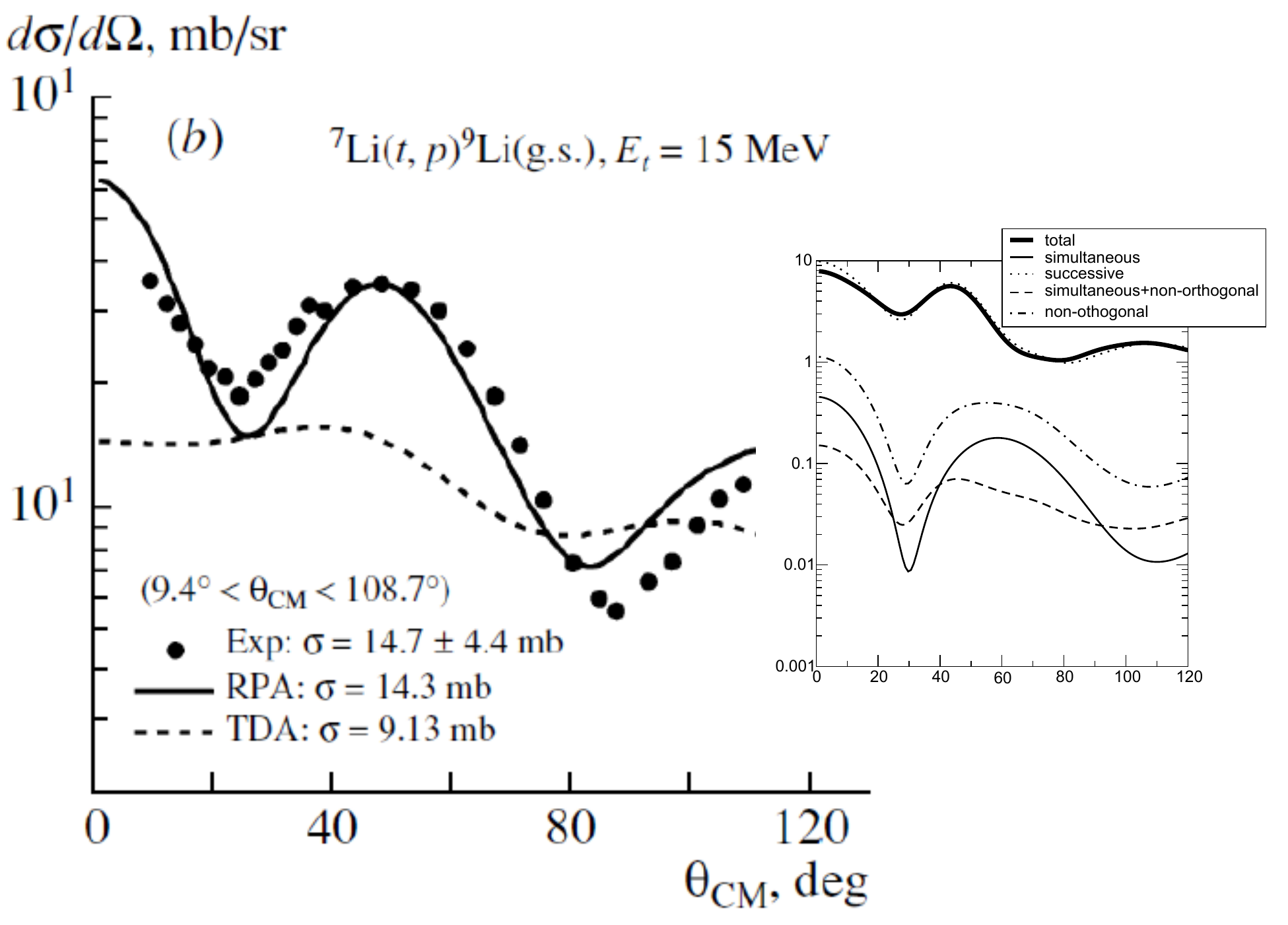}
\caption{Absolute differential cross section associated with the reaction $^7$Li(t,p)$^9$Li(gs) calculated making use of the forwardsgoing and backwards going amplitudes displayed in Fig. 2. The dashed curve corresponds to the result obtained by neglecting the backwards going amplitudes, normalising the X's to 1 (TD approximation). In the inset the variety of contributions (successive, simultaneous, non-orthogonality) to the  cross section are shown.  }
\end{center}
\end{figure}

It is quite suggestive the presence in $^{12}$Be, of a quadrupole pair addition mode almost degenerate 
with the halo monopole pair addition mode $0^{+*}$. One can thus  expect 
important  quadrupole dynamic deformation effects  resulting  from this degeneracy.
% the almost degeneracy of these states 
%(feature at the basis of the description of they interplay between pairing and  quadrupole deformation in the IBM \cite{Iachello1987})
% (within this context see also \cite{Broglia1981,Bohr82}, cf. also \cite{Hamamoto}. 
 Within this context, parity inversion arises because of Pauli repulsion between the $p_{1/2}$ nucleon in $^{10}$Li ($^{11}$Be) 
 and that participating in the quadrupole vibration of the core ($^9$Li, $^{10}$Be). The polarization self energy processes make the 
 $s_{1/2}$ particle heavier and thus closer to becoming bound  \cite{EPJ_2001,Gori}, see also \cite{Hamamoto}.  
 
 The fact that one is now able to accurately calculate two-nucleon transfer absolute differential cross sections \cite{Potel2013}  opens 
a number of possibilities, in particular to find new elementary modes of excitation in exotic nuclei.
A simple, but nonetheless instructive example of the consistency of the physics and associated accuracy of the results 
which is at the basis of clothed, physical elementary modes of excitation as building blocks of the nuclear spectrum, 
is provided by the $^7$Li(t,p)$^9$Li (gs) absolute differential cross section. As seen from Figs. 1 and 3, theory provides 
an accurate account of the experimental findings \cite{Potel2014,Young1971}.
The two-nucleon spectroscopic amplitudes were calculated  by solving   the $\alpha = -2$ 
monopole dispersion relation \cite{BesandBroglia1966,Potel2014}  in the RPA. The results are shown in Fig. 2. Eliminating ground state correlations theory underpredicts experiment by about 
50\% (cf. Fig. 3). In other words, even the ground state of an apparent "normal" nucleus 
like $^7$Li ($S_{2n} $= 12.91 MeV), resents of the properties displayed by the exotic nucleus $^{11}$Li(gs). In fact, the population of the 
pair removal mode  through ground state correlation proceeds 
by the pick-up of  $s,p$ parity-inverted orbits, typical of the neutron halo pair addition mode. 

%Within this context it is intriguing the insight that $^{10}$Be(p,t)$^8$Be(gs) and $^8$Be(t,p)$^{10}$Be can provide within the 
%study of pair addition and removal modes at  the neutron drip line.

Within this context one expects that much insight on the interplay  between the GPDR and the  monopole  neutron halo pair addition modes emerges from the systematic study 
of the reactions   \hspace{4cm}  $^{10}$Be(p,t)$^8$Be(gs), $^8$Be(t,p)$^{10}$Be, $^{10}$Be(t,p)$^{12}$Be,  $^{14}$Be(p,t)$^{12}$Be as well as those associated with 
(p,p2n) knockout reactions and eventually  2n-transfer induced by heavy ions (e.g. $^{18}$O,$^{16}$O) (Fig. 1).  An important example of such insight 
is provided by the fact that while the cross sections associated with the ground state and two-phonon (normal) monopole pairing 
vibrational states ($E_x \approx $ 4.8 MeV in $^{10}$Be), i.e.  \\ $d\sigma (^8$Be(t,p)$^{10}$Be(gs))/$d\Omega$ and
$d\sigma (^8$Be(t,p)$^{10}$Be($0^+$; 4.8 {\rm MeV} ))/$d\Omega$ 
  are expected to have the same order of magnitude (cf. Fig. 13 of \cite{Potel2014}),
  that associated with the $0^{+*}$ state  in $^{12}$Be  is predicted to be  much smaller (observable?), reflecting the poor overlap between halo and core nucleons \cite{Fortune1994} (within this context see Table 3 of ref. \cite{Potel2014} and associated discussion).   
  
 Arguably,  one would be able to state that a real understanding of the neutron halo pair addition pattern displayed in Fig. 1 has been obtained, 
once the two-nucleon transfer predictions are tested, supplemented with one-particle and $\gamma-$decay data,  worked out making use of microscopically calculated optical (polarization) potentials,
with the help of the same physical modes to be probed.


\begin{thebibliography}{99}
 \bibitem{Mottelson} B.R. Mottelson, {\it Elementary features of nuclear structure},  in  Trends in nuclear physics, 100 years later, Proc. of Les Houches summer school on theoretical physics, Session LXVI, 
 eds.  H. Nifenecker, J.-P. Blaizot, G. F. Bertsch, W. Weise, F. David, Elsevier (Amsterdam), p. 25 (1998)
 \bibitem{Brogliaetal1971c} R.A. Broglia, V. Paar and D.R. B\`es, {\it Diagramatic perturbation treatment  of the effective
interaction between two-phonon states in closed shell nuclei: the $J^{\pi} = 0^+$ states in $^{208}$Pb}, Phys. Lett. B {\bf 37} (1971) 159
\bibitem{Bohr1964} A. Bohr, {\it Elementary modes of nuclear excitation and their coupling},
in Comptes Rendus du Congr\`es International de Physique Nucl\`eaire, Vol. I,  P. Gugenberger ed., Editions du Centre Nationale de la Recherche Scientifique,
Paris (1964), p. 487 
\bibitem{BohrMottelsonII} A. Bohr and B.R. Mottelson, {\it Nuclear structure}, Vol. II, Benjamin, New York  (1975) 
 \bibitem{BesandBroglia1966} D. R. Bes and R.A.Broglia,
{\it Pairing vibrations}, Nucl. Phys. {\bf 80}, 289 (1966)
\bibitem{BohrMottelsonI}A. Bohr and B.R. Mottelson,  {\it Nuclear structure}, Vol. I , Benjamin, New York (1969)
\bibitem{Bes1976a}   D. R. B\`es, R.A.Broglia, G. G. Dussel, R. J. Liotta and H. M. Sofia,
{\it The nuclear field treatment of some exactly soluble models}, Nucl. Phys. 
{\bf A260}, 1 (1976)
\bibitem{Bes1976b}  D. R. B\`es, R.A.Broglia, G. G. Dussel, R. J. Liotta and H. M. Sofia,
{\it Application of the nuclear field theory to monopole interactions which 
include all the vertices of a general force}, Nucl. Phys. 
{\bf A260}, 27 (1976)
\bibitem{Bes1976c} D. R. B\`es, R.A.Broglia, G. G. Dussel, R. J. Liotta and R. J. Perazzo,\\
{\it On the many-body foundation of the nuclear field theory}, Nucl. Phys.  {\bf A260}, 77 (1976)
 \bibitem{Bortignonetal1977} P. F. Bortignon, R.A.Broglia, D. R. B\`es and R. Liotta,
{\it Nuclear field theory}, Phys. Rep. {\bf 30C}, 305 (1977)
\bibitem{Broglia1999} R.A. Broglia and A. Winther, {\it Heavy ion reactions}, Addison-Wesley, Menlo Park (1999)
\bibitem{Broglia1973} R.A. Broglia, O. Hansen and C. Riedel, {\it Two-neutron  transfer reactions and the pairing model}, Adv. Nucl. Phys. {\bf  6} (1073) 287
(see www.mi.infn.it/~vigezzi/BHR/BrogliaHansenRiedel.pdf)
\bibitem{Potel2013}  G. Potel, A. Idini, F. Barranco, E. Vigezzi and R.A. Broglia,  {\it Cooper pair transfer in nuclei}, Rep. Prog. Phys. {\bf 76}, 106301 (2013)
\bibitem{EPJ}  F. Barranco, P.F.Bortignon, R.A. Broglia, G. Col\`o and E. Vigezzi,  {\it The halo of the exotic nucleus $^{11}$Li: a single Cooper pair}  Eur. Phys.  J. A{\bf 11}, 385 (2001)
%\bibitem{Lenske} S.E.A. Orrigo and H. Lenske, {\it Pairing resonances and continuum spectroscopy of $^{10}$Li}, Phys. Lett. B {\bf 677}, 294 (2009)
\bibitem{Sagawa} H. Sagawa, B.A. Brown and H. Esbensen,  {\it Parity inversion in the N=7 isotones and the  pairing blocking effect}, Phys. Lett. B {\bf 309} (1993) 1
\bibitem{Tanihata2008} I. Tanihata et al., {\it  Measurement of the two-halo neutron  transfer reaction $^1H(^{11}Li,^9Li)^3H$ at 3 MeV}, Phys. Rev. Lett. {\bf 100} (2008) 192502
\bibitem{Potel2010}G. Potel, F.  Barranco, E. Vigezzi and R.A. Broglia, {\it Evidence for phonon mediated pairing interaction
in the halo nucleus $^{11}$Li}, Phys.. Rev. Lett. {\bf 105} (2010) 172502
\bibitem{Potel2014} G. Potel, A. Idini, F. Barranco, E. Vigezzi and R.A. Broglia, {\it Nuclear field theory predictions
for $^{11}$Li and $^{12}$Be: shedding light on the origin of pairing in nuclei}, Phys. At. Nucl. {\bf 77} (2014) 941  
\bibitem{Broglia1971a} R.A. Broglia, C. Riedel and T. Udagawa,  {\it  Coherence properties of two-neutron transfer reactions and their 
relation to inelastic scattering}, Nucl. Phys. A {\bf 169} 225 (1971) 
%\bibitem  {Iachello1987} F. Iachello and A. Arima, {\it The interacting boson model}, Cambridge University Press, Cambridge (1987) 
%\bibitem{Bohr82} A. Bohr and B.R. Mottelson, {\it On the ability of  the interacting boson model to describe nuclear-deformation 
%effects},  Phys. Scripta {\bf 25} (1982) 915
%\bibitem{Broglia1981} R.A. Broglia, {\it    Microscopic structure of the intrinsic state of deformed nuclei}, in 
%  Interacting Bose-Fermi systems in nuclei, F. Iachello Ed., Plenum Press, New York (1981), p.95
\bibitem{EPJ_2001}  F. Barranco,  P.F. Bortignon, R.A. Broglia, G. Col\`o and E. Vigezzi,  
{\it The halo of the exotic nucleus $^{11}$Li: a single  Cooper pair,} Eur. Phys. J. A {\bf 11}, 305  (2001)
\bibitem{Gori} G. Gori, F. Barranco,  E. Vigezzi and R.A. Broglia, {\it Parity inversion and breakdown of shell closure in Be isotopes},
Phys. Rev. C {\bf 69}, 041302(R) (2004)
\bibitem{Hamamoto} I. Hamamoto and S. Shimoura,  {\it   Properties of $^{12}$Be and $^{11}$Be in terms of single-particle motion in deformed potential
}, J. Phys. G {\bf 34}, 2715  (2007)
\bibitem{Young1971} P.G. Young and R.H. Stokes, {\it New states in $^9$Li from the reaction $^7Li(t,p)^9Li$}, Phys. Rev.
C {\bf 4} (1971) 1597
 \bibitem{Fortune1994} H. T. Fortune, G.B. Liu and D.E. Alburger, {\it $(sd)^2$ states in $^{12}$Be}, Phys. Rev. C {\bf 50} (1994) 1355

\end{thebibliography}
\end{document}